\documentclass[prl, twocolumn, a4paper, showpacs]{revtex4}
\usepackage{amssymb}
\usepackage{amsmath}
\usepackage{graphicx}
\begin{document}
\title{$\pi$-junction to probe antiphase s-wave pairing in iron pnictide superconductors}
\author{Wei-Qiang Chen$^{1}$, Fengjie Ma$^{2,3}$, Zhong-Yi Lu$^{2}$, and Fu-Chun Zhang$^{1}$} 
\affiliation{$^{1}$Department of Physics, and Center of Theoretical and Computational
Physics, the University of Hong Kong, Hong Kong, China\\
$^2$Department of Physics, Renmin University of China, Beijing 100872, China\\
$^{3}$Institute of Theoretical Physics, Chinese Academy of Sciences, Beijing 100190, China\\}
\date{\today}

\begin{abstract}
Josephson junctions between a $FeAs$-based  superconductor with antiphase 
s-wave pairing and a conventional s-wave superconductor are studied. The translational invariance in
a planar junction between a single crystal pnictide and an aluminum metal greatly enhances the 
relative weight of electron pockets in the pnictide to the critical current. 
In a wide doping region of the pnictide, a planar and a point contact junctions have opposite phases, which 
can be used to design a tri-junction ring with $\pi$ phase to probe the antiphase pairing. 
\end{abstract}


\maketitle
One of important issues for the newly discovered iron pnictide superconductors (SCs)
\cite{kamihara_iron-based_2008, chen_superconductivity_2008, chen_superconductivity_2008-1,
  ren_superconductivity_2008, wen_superconductivity_2008,
  rotter_superconductivity_2008,  wang_thorium-dopinginduced_2008, hsu_superconductivity_2008, hhwen2}  
is their pairing symmetries.
A number of experiments \cite{chen_bcs-like_2008, ding_observation_2008,zhang_observation_2008,Tsuei} have
suggested a spin singlet s-wave pairing. The iron pnictide 
has  hole and electron Fermi pockets (Fig. 1).
Theories have predicted an antiphase $s$-wave or $s_{\pm}$ state, where the pairing 
has an s-wave symmetry, but
the order parameters on the electron and hole
pockets have opposite signs 
\cite{mazin_unconventional_2008, kuroki_unconventional_2008, Tesanovic, 
seo_pairing_2008, wang_functional_2009, 
chen_strong_2009}. 
It will be important to confirm the $s_{\pm}$ state for pnictides, especially using 
more decisive phase sensitive experiments, which provided 
a direct evidence for the $d_{x^2-y^2}$ pairing for 
cuprates\cite{kirtley_symmetry_1995, wollman_experimental_1993}. 
Since $s_{\pm}$ phase is
related to the $\vec k$-space location instead of the orientation,
different types of phase sensitive experiments are needed to probe 
the signs of the gap functions\cite{tkng,tsai_novel_2008,parker_possible_2008,philip,Linder}.  
Very recently,
Chen et al.\cite{Tsuei} have carried out a new phase sensitive experiment on
polycrystal $NdFeAsO$ compound, and observed integer and half integer flux quantum transitions 
in a niobium/Fe-pnictide loop, which clearly
demonstrates the sign changes of the order parameters in Fe-pnictides. 

In this Letter, we study Josephson junctions between a $FeAs$-based  SC of 
$s_{\pm}$-pairing and a conventional s-wave SC. The translational invariance in
a planar junction between a single crystal pnictide and an aluminum metal 
greatly enhances relative weight of electron pockets to the critical current. 
In a wide doping region for both $BaFe_2As_2$ ($122$ hereafter) and $LaFeAsO$ ($1111$ hereafter) compounds, 
a planar and a point contact junctions have opposite phases. 
This property can be used to 
 design a tri-junction ring with $\pi$ phase to probe the antiphase s-wave pairing.

\begin{figure}[htbp]
\centerline{\includegraphics[width=0.2\textwidth]{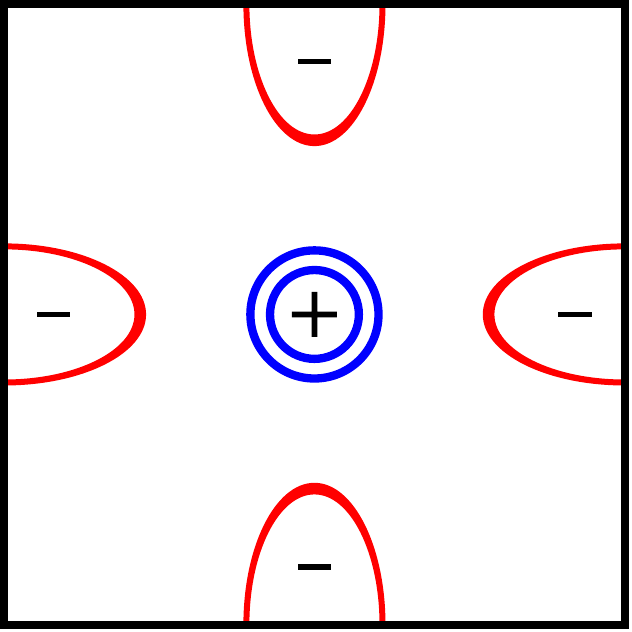}}
\caption[]{\label{fig:fs} Schematic plot of Fermi surface of pnictide in an extended Brillouin zone where each unit
  cell contains one Fe-ion.  Signs represent phase of the superconducting gap.}
\end{figure}

We start with a brief review on the charge current $I_J$ passing through a Josephson junction of two
conventional SCs.   $I_J
= I_c \sin \delta \phi$, with $I_c$ the critical current, and $\delta \phi$ the phase difference
between the two SCs. In the absence of magnetic fields or magnetic impurities,
we may focus on $I_c$, which is of the form
\begin{align}
\label{eq:1}
I_c &\propto \int d \mathbf{k} d \mathbf{q} ~ \frac{|T_{\mathbf{k} \mathbf{q}}|^2
  \Delta_1(\mathbf{k})\Delta_2(\mathbf{q}) }{E_1(\mathbf{k}) E_2(\mathbf{q}) [E_1(\mathbf{k}) + E_2(\mathbf{q})] },
\end{align}
where $T_{\mathbf{k} \mathbf{q}}$ is the tunneling matrix, $E_{i}(\mathbf{k}) = \sqrt{\epsilon_{i}
(\mathbf{k})^2 + \Delta_{i}(\mathbf{k})^2}$ is the quasiparticle energy of the SC $i=1,2$,
$\epsilon_{i}(\mathbf{k})$ is the single electron energy measured relative to the chemical potential,
and $\Delta_{1(2)}$ are superconducting gap functions (assumed to be real). 
The proportional coefficient in $I_c$ are always positive throughout this paper. 
The junction is a $0$-junction if $I_c >0 $, and a $\pi$-junction if $I_c < 0$.
If both SCs are simple s-wave, $\Delta_i(\mathbf{k})$ are independent of $\mathbf{k}$, and
 the sign of $I_c$ is determined by the relative signs of
$\Delta_1$ and $\Delta_2$.  Though $\Delta$ is not gauge invariant,
Sigrist and Rice\cite{sigrist_paramagnetic_1992} have
pointed out that the $\pi$-junctions can not be gauged away if there are odd numbers of 
$\pi$-junctions in a loop. 

Eq. (1) can be extended to study Josephson junctions between a pnictide
and a conventional SC. Let $\Delta_2$ be the superconducting gap and $E_2(\mathbf{q})$ the quasi-particle energy
of a conventional s-wave SC, and
$\gamma$ the band index of the pnictide.  
We shall neglect the inter-band pairing amplitude in the pnictide, whose effect is small due to the 
energy splitting of the bands~\cite{chen_strong_2009}. 
Denoting $\Delta_1^{\gamma}(\mathbf {k})$ the superconducting gap of the band $\gamma$ and
$E_1^{\gamma}(\mathbf{ k})$ the corresponding quasi-particle energy, we have 
\begin{align}
\label{eq:777}
I_c \propto \Delta_2 \sum_{\gamma} \int d\mathbf{k} d\mathbf{q} 
\frac{|T_{\mathbf{k} \mathbf{q}}|^2 \Delta_1^\gamma (\mathbf{k})}{E_1^\gamma(\mathbf{k})
  E_2(\mathbf{q}) [E_1^\gamma(\mathbf{k}) + E_2(\mathbf{q})]}.
\end{align}
Because of small value in $\Delta/E$ for states far from the Fermi surface, the integral 
in Eq. (2) is of appreciable value only for $\mathbf {k}$ and $\mathbf {q}$ near their Fermi surfaces.  
The Fermi pockets in Fe-pnictide are well separated, so we may replace $\sum_{\gamma} \int d\mathbf {k}$  
in Eq. (2) by integrals over $\mathbf{k}$ within a small cut-off around each Fermi pocket 
$\alpha$.  Assuming  $\Delta_1^{\alpha}$ to be isotropic 
near each Fermi pocket $\alpha$, we have
\begin{align}
\label{eq:7}
I_c \propto \Delta_2 \sum_{\alpha} \Delta_1^{\alpha}
\int d\mathbf{k} d\mathbf{q} 
\frac{|T_{\mathbf{k} \mathbf{q}}|^2 }{E_1^\alpha(\mathbf{k})
  E_2(\mathbf{q}) [E_1^\alpha(\mathbf{k}) + E_2(\mathbf{q})]},
\end{align} 
where the sum of $\alpha$ is over all the Fermi pockets within the Brillouin zone (BZ), 
and integral of $\mathbf {k}$
is over around the Fermi pocket $\alpha$ within a small cut-off. 

We choose a convenient gauge where 
the gap function of the conventional
s-wave SC is positive and the gap function of the hole (electron) pockets in the 
Fe-pnictide SC is positive (negative).
To simplify the calculations, we shall neglect the dispersion of the Fermi surface in the direction 
perpendicular to the Fe-plane in the pnictide
\cite{ding_observation_2008}.  
In what follows we will use Eq. (3) to study point and planar junctions. While our formalisms may be applied to 
all the iron-based SC, we will primarily discuss $122$-based SC for its availability of good single crystals.  
The junctions for $1111$-based SC will be briefly discussed.

\begin{figure}[htbp]
\centerline{\includegraphics[width=0.3\textwidth]{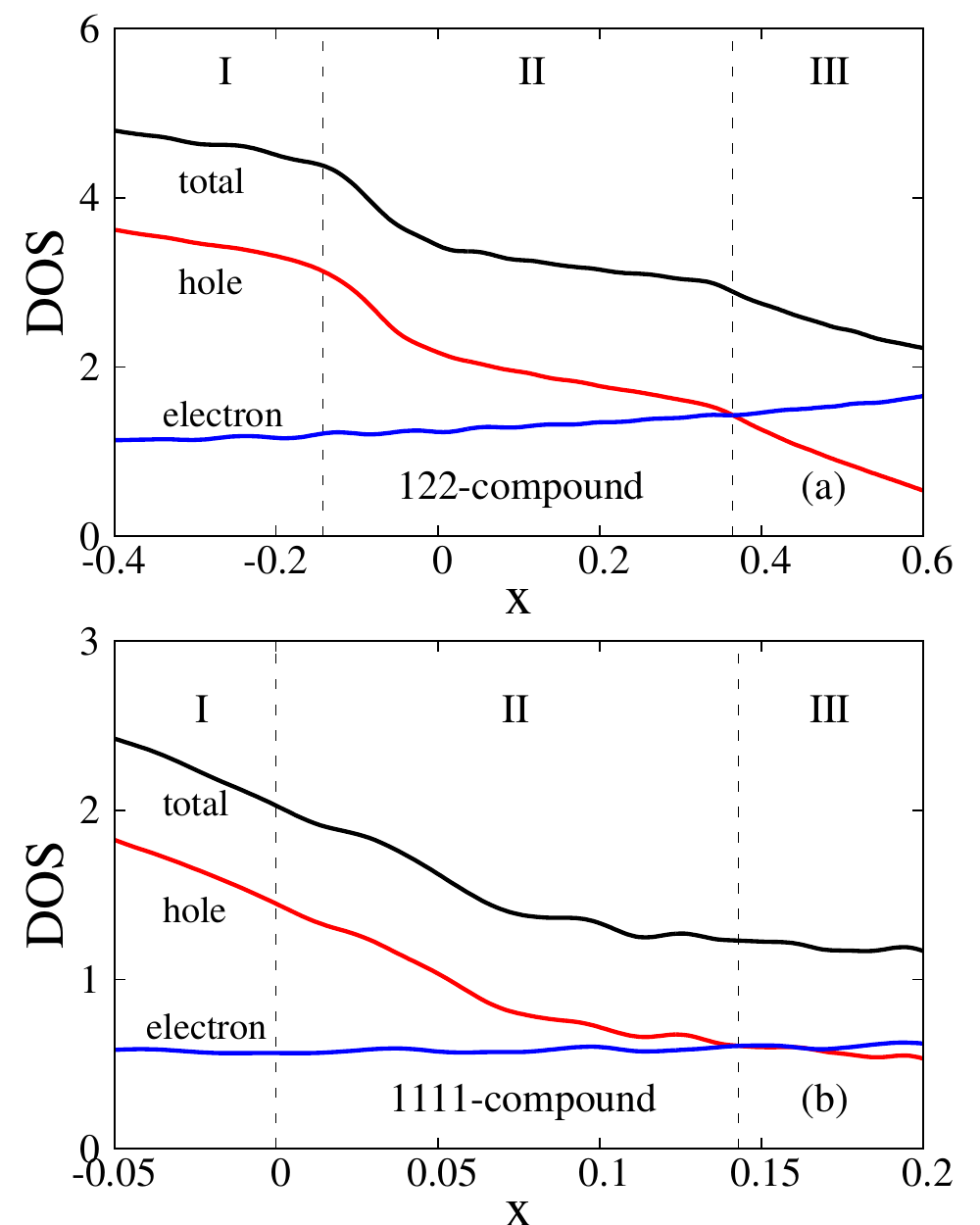}}
\caption[]{\label{fig:DOS} (color online) Hole DOS $N_h$ (red curve) and electron DOS $N_e$ ( blue curve), and total DOS
  vs doping $x$ for $122$-compound (a) and $1111$-compound (b), where $x>0$ for electron and $x<0$ for hole dopings.
  The point junction is a $0$-junction in Region I and II, and a $\pi$-junction in Region III.  The planar junction is a
  $0$-junction in Region I, and a $\pi$-junction in Region II and III.  So the tri-junction loop in Fig. 4 has a
  $\pi$-phase in Region II.}
\end{figure}

\textit{Point junction between $s_{\pm}$ and s-wave SC.}
For point junction, there is no momentum conservation in the tunneling. 
The tunneling direction relative to the FeAs plane is also random, 
so that the tunneling matrix is insensitive to the d-orbitals hence to the bands in the Fermi pockets.
We may then set $T_{\mathbf{k} \mathbf{q}}=T_0$ to be a constant, and Eq. (3) becomes
\begin{align}
I_c &\propto |T_0|^2 \sum_{\alpha} \Delta^{\alpha}_1 \Delta_2 N_{1F}^{\alpha} N_{2F}
\int d \epsilon_1 d \epsilon_2 ~ \frac{1 }{E_1^{\alpha} E_2 [E_1^{\alpha} + E_2] }, \nonumber
\end{align}
where 
$N_{1F}^{\alpha}$ and $N_{2F}$ are the density of states (DOS) on the Fermi pocket $\alpha$ of the pnictide and 
of the s-wave SC at the Fermi level, respectively.  
Following Ambegaokar and Baratoff
\cite{ambegaokar_tunneling_1963}, assuming $\Delta_2 << |\Delta_1^{\alpha}|$, we have
\begin{align}
\label{eq:3}
I_c &\propto N_{2F} \Delta_2 \sum_{\alpha} \mathrm{sgn}(\Delta_1^{\alpha}) N_{1F}^{\alpha} K[\sqrt{ 1 -
  \frac{\Delta_2^2}{\left( \Delta^{\alpha}_1 \right)^2} }],
\end{align}
where $K(x)$ is the first kind complete elliptic integral. The amplitude of $\Delta^{\alpha}_1$ of the Fe-pnictide are of the
same order \cite{ding_observation_2008}. In the limit of $\left| 4
\Delta_1^{\alpha} / \Delta_2 \right| \gg 1$, $I_c$ is given by
\begin{align}
\label{eq:4}
I_c \propto \sum_{\alpha \in BZ} \mathrm{sgn}(\Delta_1^{\alpha}) N_{1F}^{\alpha}.
\end{align}
where the sum is over pockets $\alpha$ within the BZ.
Therefore, $I_c >0$ if the hole DOS $N_h$ is larger than the electron DOS $N_e$, and
$I_c <0$ if $N_h < N_e$.  We have carried out density functional theory (DFT) calculations and the
results for $122$-compounds are shown in Fig.~\ref{fig:DOS}(a). 
From the figure, the point contact junctions are 
$0-$junctions except at very large electron doping.

\begin{figure}[htbp]
\centerline{\includegraphics[width=0.4\textwidth]{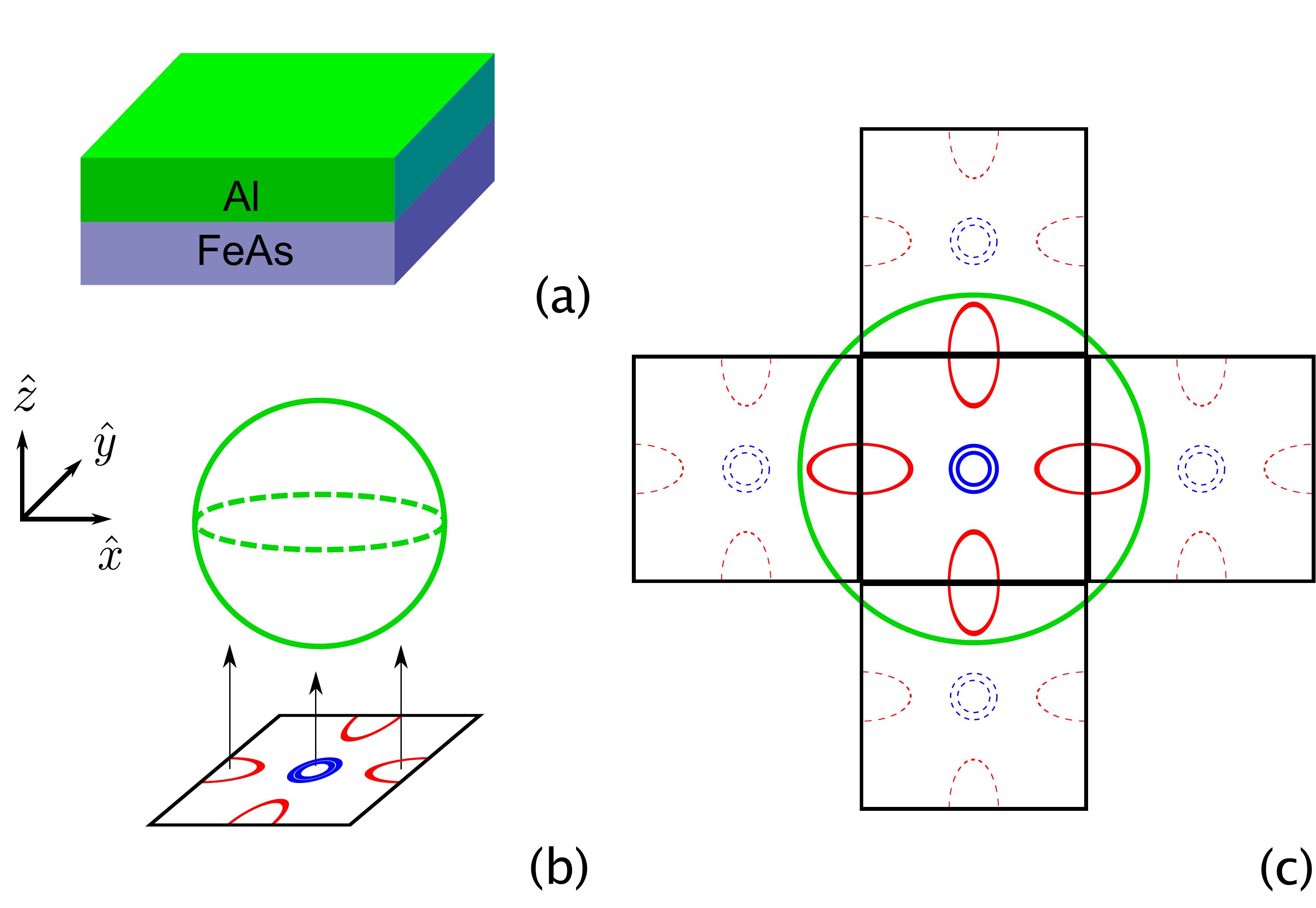}}
\caption[]{\label{fig:planar} (color on line) (a) Proposed planar junction, whose interface is parallel to the
$x-y$-plane.  (b) Illustration of electron tunneling (black arrows) with planar momentum conservation.
Top sphere  (green): Fermi surface of Al metal;
bottom:  Fermi surface of Fe-pnictide.  (c) Fermi
surface of Al (green line) and Fe-pnictide (blue lines for hole and red
for electron pockets) at $q_z = 0$. Black lines are the BZ
boundary.  The Fermi pockets outside the
green circle has no contribution to the Josephson current.}
\end{figure}

\textit{Planar junction.}  We now consider a junction between a
$122$ single crystal and a nearly free electron 
Al with a spherical Fermi surface, as shown in Fig.~\ref{fig:planar}. 
The thin insulator
plane in the junction is parallel to the FeAs-plane ($x-y$ plane) of
the pnictide.  In a nearly free electron metal, the lattice
potential of the material is very weak, so that we may neglect the
lattice effect, and the junction has a translational symmetry in the
$x-y$ plane.  Otherwise, we assume the tunneling matrix element to be independent 
of the Fermi pockets, 
$T_{\mathbf{k} \mathbf{q}} = T_0 \delta_{k_{xy},
q_{xy}}$, where $k_{xy}$ and $q_{xy}$ are the planar wavevectors.  
This assumption is valid if there are only two orbitals $d_{xz}$ and $d_{yz}$ are involved, 
for they are related by a $90^0$ rotational symmetry. 
In Fe-pnictide, there are five $d-$ orbitals with more weight on $d_{xz}$ and $d_{yz}$, but finite weight on others
\cite{2orbital}. We expect that our approximation be reasonably good, and will return to discuss
the correction to this approximation. 

It is convenient to work on the repeated zone scheme, 
where the BZ of the wavevector $\mathbf {k}$ of the pnictide 
is expanded into an infinite plane. Because of the planar momentum conservation, noting that
the integral in Eq. (2) is only of appreciable value near the Fermi surfaces/pockets, 
Eq. \eqref{eq:7} becomes,

\begin{align}
\label{eq:5}
I_c &\propto |T_0|^2 \Delta_2 \int\limits_{|k_{xy}| < q_F } dk_{xy} dq_z \sum_{\alpha}
\frac{\Delta^{\alpha}_1}{E^{\alpha}_1(k_{xy}) E_2(k_{xy}, q_z)}
\nonumber\\
& \phantom{|T_0|^2 \Delta_2 \sum_{\alpha \in q_F} \Delta^{\alpha}_1} \times \frac{1}
{E^{\alpha}_1(k_{xy}) +
  E_2(k_{xy}, q_z)}
\end{align}
where $q_F$ is the Fermi wavevector and $q_z$ the z-component wavevector 
of the conventional SC. The 
sum of $k_{xy}$ is within the circle of a radius $q_F$ in the x-y plane, and 
$\sum_{\alpha}$ stands for the summation over all the Fermi pockets within the 
circle of radius $q_F$ as illustrated in Fig. 3.  The tunneling process for the wavevector outside the BZ
is due to the umklapp process.

For a free electron metal, we have $\int d q_z \approx
\frac{C}{\sqrt{q_F^2 - k_{xy}^2}} \int d \epsilon_2$, with $C$ a
constant.  To further calculate $I_c$, we approximate
$k_{xy}$ in the term of $1/\sqrt{q_F^2 - k_{xy}^2}$ by a mean
squared average of the wavevector $Q_{\alpha}$ within the Fermi
pocket $\alpha$, and obtain 
\begin{align}
\label{eq:6}
I_c &\propto T_0^2 \Delta_2 \sum_{\alpha \in q_F} \frac{\mathrm{sgn}(\Delta_1^{\alpha}) 
N_{1F}^{\alpha}}{\sqrt{q_F^2 -
    Q^2_{\alpha}}} K(\sqrt{ 1 - \frac{\Delta_2^2}{\left( \Delta^{\alpha}_1 \right)^2} })
\nonumber\\
&\sim \Delta_2 \sum_{\alpha} \frac{\mathrm{sgn}(\Delta_1^{\alpha}) 
N_{1F}^{\alpha}}{\sqrt{q_F^2 - Q^2_{\alpha}}}
\end{align}
For Al, the Fermi surface is a sphere (see Fig.~\ref{fig:planar}(b))
of a radius of the Fermi wavevector $q_F= 1.75 \AA^{-1} = 1.56
\pi/a$, with $a =2.8 \AA$ the distance of two nearest neighbor irons
in 122 compound.  As we can see from Fig.~\ref{fig:planar}(c),
the area hence the DOS of the  electron pockets enclosed within a
circle of $q_F$ is twice of the area or the DOS within the BZ
 of the Fe-pnictide.  By using Eq. (7) and the results of $N_e$ and $N_h$ from DFT,
approximating $Q_{\alpha} =\pi/a$ for the electron pockets, and 
 $Q_{\alpha}=k_F^h(k_z=0)$, the Fermi momentum for the hole pockets, we 
 find $I_c <0$ when $N_h / N_e < 2.58$ which corresponds to the region II and III in Fig. 2(a). In our calculations, 
we have assumed the tunneling matrix to be independent of the momentum or the
Fermi-pockets. In 122 compounds, in addition to $d_{xz}$ and $d_{yz}$ orbitals, 
the electron configuration on the hole pockets contains $d_{3z^2-1}$ orbital, 
and also contains more $d_{x^2-y^2}$ or $d_{xy}$ orbitals than on the electron pockets. Because 
of their orbital orientations, the $d_{3z^2-1}$ orbital enhances the tunneling matrix, and the
$d_{x^2-y^2}$ and $d_{xy}$ orbitals do the opposite.  Therefore, their overall effect to 
the condition of the $\pi$-junction is partially canceled. 
We expect the results obtained by this approximation to be qualitative or semi-qualitatively correct.

We have also combined DFT results and ARPES data to calculate $I_c$ and obtained similar results.
In the calculation, we use Eq.~\eqref{eq:7} directly by taking into account of the planar momentum conservation.  
We apply a rigid band approximation for the normal state electron state in DFT and take
the gap functions of the hole doped 122 compound
from ARPES data of Ding et al. The dispersion in Al is modeled
by $\epsilon_{\mathbf{q}} = q^2 \hbar^2/2m^{*}$, with the effective mass of the electron $m^{*}=1.16 m_e$, and $m_e$
the free electron mass.   We assume a BCS gap
function for Al SC  at T=0 based on value of $T_c =1.175 K$. We have found the planar junction  is of 
$\pi$-phase at hole doping up to 0.4, which gives a wider region for the $\pi$-junction. 

\begin{figure}[htbp]
\centerline{\includegraphics[width=0.35\textwidth]{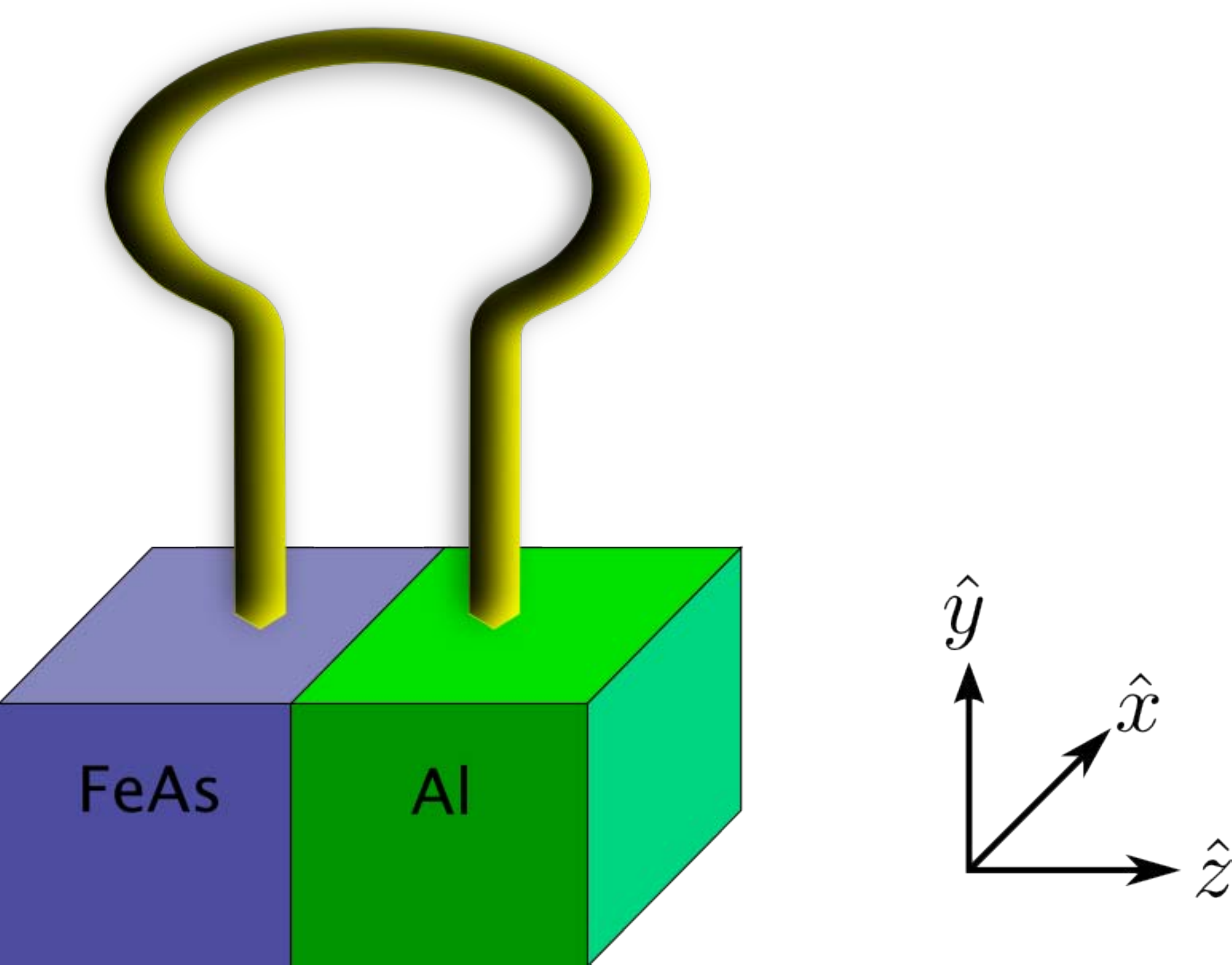}}
\caption[]{\label{fig:setup} A Tri-junction setup, consisting of a planar junction between Al and Fe-pnictide
  SC, and a point junction between a conventional SC (say niobium) and Al and a point junction between the niobium and
the pnicitide.}
\end{figure}

\textit{Tri-junction to probe s$_{\pm}$ symmetry }
We now discuss Josephson tri-junction to probe the $s_{\pm}$-pairing in Fe-pnictide SC. 
The experimental setup is illustrated in Fig.~\ref{fig:setup}, 
similar to that in Ref. \cite{Tsuei}.
Fe-pnictide is chosen so that
its planar junction with Al is a $\pi$-junction, and its point junction with niobium is a $0$-junction.
The tri-junction configuration is then a $\pi$-loop characterized by 
half-integer flux quantization.  The condition for the such a $\pi$-junction of the planar and
0-junction of the point junctions is illustrated in Fig. \ref{fig:DOS}(a)  for 122-compound. 
As we can see, there is a large region of the material space
to satisfy the $\pi$ tri-junction condition.

Let us briefly discuss the electron doped 1111-compound, which will be important 
when its single crystal becomes available.  Our DFT calculations show that $N_h/N_e >1$ at the doping $x < 0.15$, and 
$N_h/N_e <1$ at $x>0.15$. Therefore, its point junction with a conventional SC will 
be a $0$-junction at $x<0.15$ and 
a $\pi$-junction at $x>0.15$, as shown in Fig. \ref{fig:DOS}(b).  A planar junction between 1111-compound and Al SC is found to have 
 $\pi$-phase for all the electron doped region studied.  Hence a set-up in Fig. 4 for 1111 with 
$x < 0.15$ is expected to be a $\pi$ junction loop. 

We remark the implication of the experiment observation of the $\pi$-junction loop in the set-up of Fig. 4. 
A $\pi$-junction indicates the opposite phases in the point and planar junctions the FeAs SC 
is involved.  This would rule out a conventional s-wave 
pairing, or a d-wave pairing such as $d_{x^2-y^2}$ or $d_{xy}$.  The later will 
result in a vanishing critical current in a junction with a s-wave SC. 
Thus the observation of the $\pi$-junction loop should be a clear indication of 
the $s_{\pm}$ pairing state for the FeAs SC.

\textit{Tri-junction with two-pnictide SCs}
We now turn to a discussion of a Josephson tri-junction ring, consisting of two single crystal Fe-pnictides 
and one conventional SC, say niobium. The junctions between each of the Fe-pnictide and the niobium 
are both point contact, and 
the junction between the two pnictides is a planar one with the 
junction plane parallel to the FeAs planes in both the pnictides.
We request $N_h > N_e$ in the first Fe-pnictide, and $N_h < N_e$ in the second Fe-pnictide.
Therefore, one of the point junctions is $0$-junction, and the other is $\pi$-junction.
The pnictide-pnictide planar junction
is a $0-$ junction because of the $x-y$ plane crystal momentum conservation 
in the tunneling process, where the tunnelings
only occur between hole pockets or between electron pockets in the 
two SCs. The tri-junction ring thus designed should have a $\pi$-junction in nature.
The experimental challenge to design this tri-junction is related to the sample quality and the selection of the second 
Fe-pnictide where the electron DOS is larger.  The DFT calculations suggest that 1111-compound is a good candidate
for this type of tri-junctions when the single crystals become available.

In summary, we have examined the phase of the Josephson junctions between Fe-pnictide and conventional s-wave
superconductors.  The sign of a point-contact junction is positive if the hole DOS is larger than the
electron DOS in Fe-pnictide and is negative otherwise.  
In the planar junction between a single crystal
Fe-pnictide and Al, planar translational invariance in the 
tunneling enhances the contribution of electron pockets to
the critical current.  We have proposed a 
Josephson tri-junction to probe the s$_{\pm}$ symmetry in Fe-pnictide,
which appears to be accessible in experiments.

We thank C. C. Tsuei, T. M. Rice, and 
and C. C. Chi for many useful discussions.  This work is  supported in part by HKSAR RGC 
grant and by NSFC and by National Program for
Basic Research of MOST, China, and by KITPC through the Project of Knowledge Innovation Program (PKIP) 
of CAS, Grant No. KJCX2.YW.W10.

\end{document}